\documentclass[conference,compsoc]{IEEEtran}
\IEEEoverridecommandlockouts
\usepackage{cite}
\usepackage{amsmath,amssymb,amsfonts}
\usepackage{algorithmic}
\usepackage{graphicx}
\usepackage{textcomp}
\usepackage{xspace}
\usepackage{cite}
\usepackage{amsmath,amssymb,amsfonts}
\usepackage{algorithmic}
\usepackage{graphicx}
\usepackage{textcomp}
\usepackage{amssymb} 
\usepackage{multirow} 
\usepackage{amssymb} 
\usepackage{graphicx} 
\usepackage{array} 
\usepackage{booktabs}
\usepackage{caption}
\usepackage{enumitem}
\usepackage{booktabs}
\usepackage{multirow}
\usepackage{graphicx}
\usepackage{tabularx}
\usepackage{float}
\usepackage{makecell}
\usepackage{xspace}

\usepackage{lmodern}

\usepackage{xcolor,colortbl}      
\usepackage[ruled,vlined]{algorithm2e}
\SetKwInput{KwInput}{Input}
\SetKwInput{KwOutput}{Output}
\usepackage{newfloat}
\usepackage{listings}
\usepackage{soul}
\usepackage{color}
\usepackage{caption}
\usepackage{multirow}
\usepackage[
  colorlinks=true,
  linkcolor=black,
  citecolor=black,
  urlcolor=blue
]{hyperref}
\usepackage[most]{tcolorbox} 

\tcbset{ learnedbox/.style={ colback=blue!5!white, colframe=blue!50!black, title=Snapshot of Learned Rules, fonttitle=\bfseries, coltitle=white, boxrule=0.8pt, arc=3mm, left=4mm, right=4mm, top=2mm, bottom=2mm, enhanced, } }

\def\BibTeX{{\rm B\kern-.05em{\sc i\kern-.025em b}\kern-.08em
    T\kern-.1667em\lower.7ex\hbox{E}\kern-.125emX}}

\newcommand{\methodname}{AgentCyTE\xspace}

\begin{document}

\title{\methodname: Leveraging Agentic AI to Generate Cybersecurity Training \& Experimentation Scenarios}

\author{
\IEEEauthorblockN{
Ana M. Rodriguez\textsuperscript{*}, 
Jaime Acosta\textsuperscript{*}, 
Anantaa Kotal\textsuperscript{*},
and Aritran Piplai\textsuperscript{*}
}
\IEEEauthorblockA{
Dept. of Computer Science, The University of Texas at El Paso, El Paso, TX, USA\\
amrodriguez28@miners.utep.edu, \{jcacosta, apiplai, akotal\}@utep.edu
}
\thanks{\textsuperscript{*}These authors contributed equally to the work.}
}

\maketitle

\begin{abstract}
Designing realistic and adaptive networked threat scenarios remains a core challenge in cybersecurity research and training, still requiring substantial manual effort. While large language models (LLMs) show promise for automated synthesis, unconstrained generation often yields configurations that fail validation or execution. We present \methodname, a framework integrating LLM-based reasoning with deterministic, schema-constrained network emulation to generate and refine executable threat environments. Through an agentic feedback loop, \methodname observes scenario outcomes, validates correctness, and iteratively enhances realism and consistency. This hybrid approach preserves LLM flexibility while enforcing structural validity, enabling scalable, data-driven experimentation and reliable scenario generation for threat modeling and adaptive cybersecurity training. Our framework is openly accesible at: \href{https://github.com/AnantaaKotal/AgentCyTE}{\url{https://github.com/AnantaaKotal/AgentCyTE}}
\end{abstract}


\vspace{-15pt}
\section{Introduction}
\vspace{-10pt}
A cybersecurity scenario constitutes a configurable networked environment that models systems, services, and threat behaviors for controlled study. Such environments underpin experimental research aimed at analyzing attack propagation, defensive strategies, and system resilience. They also support practical training activities such as incident response exercises and Capture-the-Flag (CTF) competitions, providing reproducible settings in which defensive mechanisms and analytical tools can be evaluated under realistic network conditions. Constructing these environments, however, remains technically demanding. Existing methods rely on network emulation frameworks such as the Common Open Research Emulator (CORE)~\cite{core}, or custom virtualized infrastructures to manually define topologies, configure routing and services, and embed vulnerabilities or attack scripts. Each component must be carefully tuned and validated to ensure internal consistency and functional execution.  The absence of automated validation or adaptive generation mechanisms, limits scalability and adaptability to evolving topology and threat scenarios.

LLMs offer a promising direction for automating scenario generation through their capabilities in structured reasoning, code synthesis, and natural-language understanding. Yet, unconstrained LLM outputs remain unreliable when used to generate executable cybersecurity scenarios. The models lack awareness of system-level dependencies, network semantics, and execution constraints necessary for operational correctness. Generated configurations often violate schema requirements, produce inconsistent or incomplete topologies, and omit critical components such as routing logic or service bindings. When executed, these deficiencies manifest as invalid network states, unreachable nodes, or non-functional services. Moreover, LLMs lack an inherent mechanism for validating or refining their own outputs, yielding scenarios that may appear correct syntactically but fail operationally. These limitations underscore the need for frameworks that combine the generative flexibility of LLMs with deterministic validation and execution control. We present \methodname, a modular framework that integrates LLM-based reasoning with schema-constrained network emulation to automate the generation and refinement of cybersecurity training and experimentation scenarios. Built on the CORE backbone, \methodname translates natural-language intent into executable, standards-compliant network environments and employs an agentic feedback mechanism that iteratively improves scenario realism and functional validity through structured observation. Our contributions are as follows:
\vspace{-2pt}
\begin{itemize}[leftmargin=*, noitemsep]
\item \textbf{Schema-Constrained Scenario Specification:} A domain-specific schema encoding topology, services, vulnerabilities, routing, and traffic behaviors as machine-verifiable elements, ensuring deterministic and executable scenario generation within CORE.
\item \textbf{Agentic LLM Integration:} A feedback-driven LLM controller that generates, validates, and refines configurations based on structured error feedback, coupling generative reasoning with programmatic verification to autonomously improve realism and consistency.
\end{itemize}

The remainder of this paper details the deterministic generation system, the architecture of the agentic feedback loop, and evaluation results demonstrating \methodname’s capability to translate natural-language specifications into executable, semantically coherent cybersecurity simulations. Our framework is openly accesible at: \href{https://github.com/AnantaaKotal/AgentCyTE}{\texttt{https://github.com/AnantaaKotal/AgentCyTE}}

\section{Related Work}
The automation of Capture-the-Flag (CTF) challenge creation has gained traction as a way to scale cybersecurity training and reduce instructor workload. Existing research, such as Automatic Challenge Generation for Hands-on Cybersecurity Training \cite{benzi2022automatic}, introduced frameworks to programmatically generate individualized cybersecurity challenges using technologies like Docker to ensure secure and isolated environments. These systems promote variability and prevent solution sharing but remain limited to predefined templates and static configurations. Similarly, Thaqi et al. \cite{thaqi2024leveraging} explored AI-driven CTF optimization to provide dynamic hints and improve accessibility, while Muzsai et al. \cite{muzsai2025improving} proposed reinforcement learning over procedurally generated cryptographic CTFs to enhance reasoning capabilities in LLM agents. Together, these works highlight a growing interest in automating challenge creation but stop short of enabling fully dynamic, network-based environments.

Beyond challenge generation, network simulation and traffic synthesis have also evolved to support realistic cyber exercise environments. Frameworks such as the Network Traffic Synthesis and Simulation Framework for Cybersecurity Exercise Systems \cite{kim2024network} and Synthetic Network Traffic Data Generation \cite{ammara2024synthetic} employ advanced generative models like CTGANs to replicate authentic network traffic patterns for use in security testing and analysis. Immersive Labs uses Gen-AI to power their Crisis Simulator to create training for defenders \cite{Immersiv36:online}. Similarly, ID2T \cite{cordero2021generating} injects synthetic attacks into background traffic to produce labeled intrusion detection datasets. Although these approaches enhance the realism of network-based exercises, most remain dataset-oriented rather than integrated into configurable emulation environments, such as CORE.

CTF hosting and deployment platforms such as TryHackMe \cite{TryHackMe} and Hack The Box \cite{HackTheBox} demonstrate the effectiveness of containerized and virtualized challenge environments. Érsok et al. \cite{ersok2024improving} highlight the logistical and technical complexities involved in managing CTF infrastructures, including container orchestration, resource allocation, and continuous integration/deployment pipelines, emphasizing the need for secure and maintainable systems. While automation tools such as Chad address aspects of challenge instantiation and deployment, and frameworks like GENIND automate network topology generation, these solutions remain distinct and often disconnected from broader emulation-based systems. Our work seeks to bridge this gap by introducing a dynamic, scriptable network emulation framework that integrates automated traffic generation and vulnerability injection via URLs within multi-node CORE configurations. This approach enables faster, reproducible, and scalable CTF scenario creation, transforming what has traditionally been a static and time-intensive setup process into a more adaptive and extensible system.

\section{System Design}
\subsection{System Overview}
The goal of \methodname is to transform the creative potential of \textbf{LLMs} into reliable, executable, and pedagogically useful cybersecurity scenarios. To achieve this, the system is designed as a modular platform that combines CORE-driven emulation, schema-constrained generation, and an \textbf{agentic feedback loop}.  

\begin{figure}[ht]
    \centering
    \includegraphics[width=.8\linewidth]{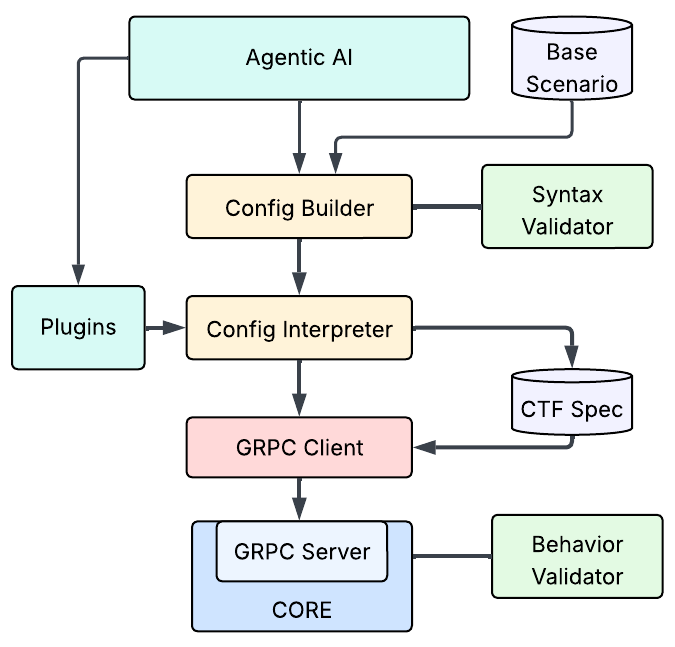}
    \caption{\methodname Workflow}
    \label{fig:placeholder}
\end{figure}

Our initial experiments confirmed our hypothesis that unconstrained LLMs struggle to produce executable scenarios. Despite extensive prompt engineering, only 38.3\% of the generated scenarios successfully executed at CORE. To address this limitation, we designed a modular system that enforces structure and reliability in scenario generation.
At a high level, \methodname (see Figure \ref{fig:placeholder}) comprises three interacting components:  

\textbf{1. Schema-Constrained Generator:}  
The generator interprets natural-language prompts and maps them onto a predefined scenario schema. The schema enforces validity across topology, routing, service configuration, and segmentation layers. The generator supports optional \textit{Base Scenarios}, such as existing CORE XML configurations that can be extended with new nodes, services, or adversarial behaviors. A built-in syntax validator ensures compliance with CORE’s XML schema prior to instantiation.  

\textbf{2. Configuration Interpreter and Emulation Backend:}  
The interpreter resolves abstract schema elements into concrete parameters (e.g., node counts, routing assignments, or service distributions) and applies a cyclic validation loop to enforce logical consistency. The backend interfaces with CORE’s gRPC API to instantiate and manage the emulated network, configure routing and services, and perform runtime verification. A behavioral validator probes connectivity between nodes, producing a matrix that verifies compliance with routing and segmentation rules.  

\textbf{3. Agentic Feedback Controller:}  
Validator logs and errors are fed back to an agentic LLM controller. The controller interprets errors and anomalies, adjusts parameters, and regenerates improved scenarios. This feedback loop enables autonomous refinement, ensuring that future generations exhibit greater functional realism and internal consistency.

\begin{figure}
    \centering
    \includegraphics[width=\linewidth]{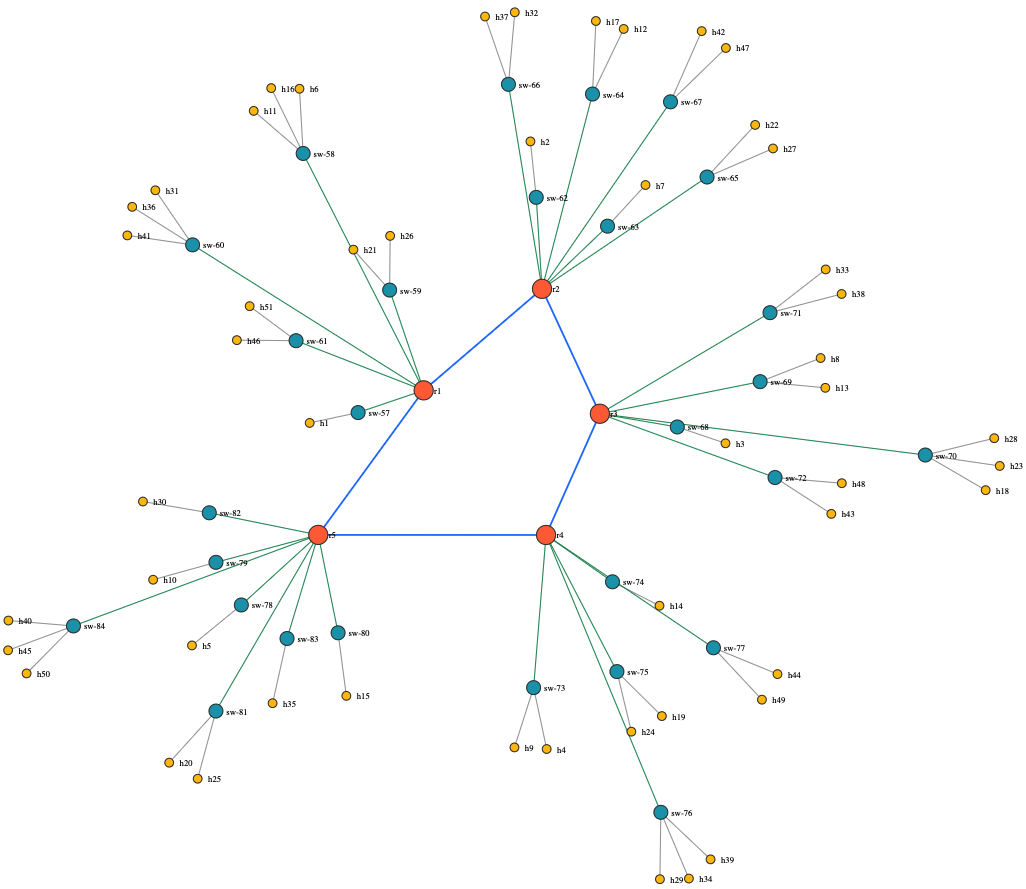}
    \caption{Topology. Orange:Rtr, Green:SW, Yellow:H}
    \label{fig:preview_window}
\end{figure}

The configuration is displayed to the user using a preview, which consolidates all information and provides a graphical view of the scenario. This information is saved to an XML file. Figure \ref{fig:preview_window} shows a sample output when 51 hosts are specified, 0.1 Router-Density, Uniform Router-to-Router edges, and Router-to-Switch from 1--3.

\noindent \textit{\textbf{- Configuration Builder and Interpreter:}} The Configuration Builder provides a way for the user to specify scenario elements. These can be defined as specific values, hints, or left to be implemented using a psuedo-random approach. The Configuration Interpreter takes the scenario elements, discussed in more detail in the following subsection, and their parameters and generates a preview with concrete topology, services, flags, etc. At this stage, a cyclic approach is used to ensure that certain constraints are met. For e.g., a router must be assigned a routing protocol or static routes; these routes must be functionally correct (meaning that e.g., if OSPFv2 is used, then surrounding routers must be configured to communicate with this service); ensuring segmentations are not redundant and logically correct (e.g., if traffic is flowing to/from nodes in certain segments, firewall rules should allow those flows). 

\noindent \textit{\textbf{- Execution Engine:}} Once approved by the user, the backend system (which may be on the local or on a remote machine) will be invoked. The backend is running a Flask/Waitress setup with an https proxy to forward requests and commands which provides a secure way to communicate with CORE's grpc interface. The system at this point will traverse through all of the elements specified in the the XML and generated associated core nodes and edges. The scenario is saved and then started. 

\noindent \textit{\textbf{- Management and Validation Interface:}} On the frontend, this data is packaged as an all-in-one zip file, which contains traffic scripts, firewall scripts, session xml, scenario xml, as well as a report that describes the scenario and flags in natural language and tabular data (everything required to re-rerun the scenario at a later time; even on different machines). Alternatively, these can be pulled individually. The frontend also provides a CORE management view, which allows a user to see all running CORE scenario instances, including a detailed view that shows a graphical representation. A user may also invoke a behavioral validator, which adds a service to each node that pings every other node. Based on the results, a connectivity matrix is then generated, which shows which nodes can communicate with others. Cross-referencing these results with the constraints (firewall rules, routers, etc.) will ensure that the scenario is functioning as expected. Stopping, restart, and removal of CORE instances can also be handled through this view.

\noindent \textit{\textbf{- Theoretical bounds on Performance:}} The performance of the various system functions are as follows.
Let $H$ denote the total number of hosts, $S$ the number of switch groups, 
$R$ the number of routers, $F$ the number of flows, and $W$ the number of weight rows. 
Table~\ref{tab:complexity} summarizes the asymptotic complexity of each step in the algorithm.

\begin{table}[h!]
\centering
\renewcommand{\arraystretch}{1} 
\begin{tabular}{|l|c|}
\hline
\textbf{Step} & \textbf{Complexity} \\
\hline
Weight normalization & $O(W)$ \\
Host assignment & $O(H)$ \\
Grouping & $O(H)$ \\
Router linking (full mesh, worst case) & $O(R^2)$ \\
Service assignment & $O(H \cdot \text{avg\_services})$ \\
Traffic flow construction & $O(F)$ \\
Segmentation rule generation & $O(S^2)$ (worst case) \\
\hline
\end{tabular}
\caption{Asymptotic complexity of each algorithm step.}
\label{tab:complexity}
\end{table}

Additionally, we designed the system to incorporate plugins, that are useful when additional capability is required. This scenario description, or CTF Specification, is saved to XML and then used to generate the live network environment in CORE.

\subsection{Scenario Specification Elements}
We wanted to make the process of creating scenarios broad or specific according to user needs. A broad description could help a user create a CTF that they could use to practice and enhance their skills, while a specific scenario could address targeted objectives for training. The following are the elements that can be tuned by a user using \methodname:

\begin{itemize}[leftmargin=-0.5pt, noitemsep]
    \item  \textbf{Base Scenario:} Defines an optional, user-provided CORE scenario (in XML format) that serves as the foundation for subsequent generation. The base scenario may include existing network artifacts such as preconfigured nodes, service instances, or hardware interfaces (e.g., RJ-45 adapter nodes for hardware-in-the-loop experimentation). During initialization, \methodname validates this scenario against a derived XML schema extracted from CORE’s source specification to ensure syntactic and semantic compatibility. Preserving base components allows the system to augment or extend existing environments rather than overwrite them, enabling hybrid scenarios that blend user-defined infrastructure with automatically generated topologies and services.

    \item \textbf{Topology Definition:} Elements can be assigned using count-based or weight-based selections. The {Node Information} section specifies the number and type of nodes that should exist in the scenario. With a count-based approach, users specify a total number of nodes and their types, e.g., five PC, 10 Workstation, five Server, and five Random. In Random mode, one of the other types is automatically chosen. In a weight-based approach, weights (values between 0--1) define type distribution. Any number of weight-based entries may be added as long as their total weight equals one. This flexibility supports both deterministic and stochastic network configurations.
    
    \item \textbf{Routing:} Nodes are connected by routers and switches. Routing elements allow routers to be assigned multiple routing protocols; by default, the pre-packaged CORE protocols (provided by the Zebra routing suite) are selectable: RIP, RIPv2, OSPFv2, OSPFv3, BGP, etc. As with nodes, these can be weight- or count-based, and multiple may be included. For e.g., if both RIP and OSPFv2 are added, the two protocol types are distributed on routers accordingly. OSPFv2 routers connect to other OSPFv2 routers, and RIPv2 routers connect similarly. CORE’s predefined Zebra services automatically generate valid routing configurations, ensuring inter-node communication. Router-to-Router and Router-to-Switch profiles can also be defined, including Min, Uniform, Non-Uniform, Exact, and Random (one selected automatically). For Non-Uniform Router-to-Switch profiles, switch-to-host edges may be defined using minimum and maximum values.

    \item \textbf{Services:} represent pre-defined or user-defined network functions distributed across nodes. Examples include SSH, HTTP, and DHCP. As with Routing, CORE ensures that configurations for each service remain valid. For e.g., selecting HTTP enables Apache on a node, creates private directories, deploys a default site (modifiable by the user), and executes validation commands to verify successful startup. This mechanism ensures correct initialization and consistency during emulation.

    \item \textbf{Traffic:} Traffic is critical to producing realistic network activity. Users can add various traffic types, including text, photo, audio, video, and random gibberish, transmitted over TCP or UDP. Each payload follows a temporal pattern—continuous, periodic, burst, Poisson, or ramp—with an optional jitter percentage, consistent with CORE’s Multi-Generator Test Tool (MGEN). Values may also be randomized. To integrate traffic with CORE nodes, the system creates two traffic scripts (sender and receiver) implementing the defined specifications. A custom CORE service is added to every communicating node, forcing it to copy and execute its associated traffic script on boot, thus generating consistent and repeatable traffic flows.

    \item \textbf{Events and Temporal Triggers:} IIn some cases, a CTF developer, experimenter, or trainer may require time-based {Events} to occur during an exercise, such as disabling a firewall to expose a previously isolated subnet or triggering a time-based exploit. \methodname allows users to specify these events by providing a path to a script executed at a given simulation time, enabling dynamic and adaptive scenario behavior.

    \item \textbf{Vulnerabilities and Flags:} Objectives in the form of \textbf{Flags} represent the primary learning and assessment components of each scenario. In \methodname, these are implemented as \emph{Vulnerabilities}, packaged either as URLs referencing Docker Compose files or as paths to binary executables. CORE supports Docker-based nodes that instantiate the compose file at boot, auto-assign IP and MAC addresses, and execute specified startup commands. This approach leverages Docker’s advantages—resource efficiency, isolation, portability, and broad image availability. \methodname enables users to specify vulnerability sources through a structured CSV file containing name, description, path, references, type (Docker Compose or binary), startup command, and vector (local or remote). The system ingests this file to populate a vulnerability catalog, where each entry can be edited, refreshed, and status-checked. Binary vulnerabilities can exist or be absent; Docker-based ones may exist, be absent, downloaded, or pulled. Users can download or remove vulnerabilities as needed and assign them through count-based, weight-based, or random selection methods. Each vulnerability type and vector can be directly specified or randomized, enabling diverse and realistic exploit configurations.

    \item \textbf{Segmentation and Isolation:} Segmentation allows definition of subnet isolation, VPNs, and network address translation mechanisms. Implemented through iptables-based firewall rules, segmentation governs traffic flow between defined network zones. The system’s cyclical validation process ensures idempotent rule generation, preventing duplication, overlap, or excessive isolation. This functionality supports the modeling of layered defense architectures and controlled access boundaries within emulated environments.

    \item\textbf{Plugin Extensions:} Services, Traffic, Segmentation, and Flags elements may be modified or augmented using plugins. Through plugins, additional user-defined services can be added to CORE that, e.g., introduce additional technologies, such as software defined networking components, wireless-specific capabilities, and others. Other protocols can be added, including custom and non-TCP as well as other profiles, etc. For Segmentation, other segregation mechanisms, such as VPN and (again) SDN elements, can be incorporated. Additional types of flags can be introduced, including custom artifacts, virtual machines, or even connections to physical hardware.

\end{itemize}

\begin{table}[h]
\centering
\resizebox{0.7\columnwidth}{!}{%
\begin{tabular}{lc}
\hline
\textbf{Model} & \textbf{Accuracy (\%)} \\
\hline
Claude & 17.33 $\pm$ 4.92  \\
GPT-4 & 19 $\pm$ 2.94 \\
GPT-OSS (120B) & 27.66 $\pm$ 6.12  \\
\hline
\end{tabular}
}
\caption{Single-shot generation accuracy for different LLMs averaged across 3 runs. 
}
\label{tab:zeroshot}
\end{table}

\vspace{-15pt}
\subsection{Agentic Feedback Controller}
\vspace{-10pt}
Initial experiments with one-shot generation of CORE XML configurations using large language models (LLMs) revealed significant reliability limitations. Despite producing syntactically well-formed outputs, models frequently failed structural validation. This discrepancy arises because the effective schema is not fixed; it varies with the degree to which the base CORE template has been populated. Furthermore, the model exhibited low output diversity, repeating identical corrections for an error regardless of their success. To quantify this limitation, we evaluated multiple state-of-the-art models under identical prompt conditions, measuring the percentage of generations that successfully passed schema validation and executed within CORE. As shown in Table~\ref{tab:zeroshot}, single-shot generation achieves less than 45\% accuracy across models, confirming that unconstrained LLM generation is insufficient for deterministic scenario construction.

To address this, we employ an agentic workflow (orchestrated by {Agno AI} \cite{agnoai}): the model first proposes a candidate XML, inspects the validation error, and attempts a repair. In practice, this naive error$\rightarrow$fix loop proved too deterministic—CORE-side issues and repetitive model edits led to cycles that did not converge.

\vspace{2pt}
\noindent\textbf{- Hypothesis-Driven Correction:} We introduce a \textit{hypothesis agent} that breaks the determinism. Instead of immediately patching the last error, the agent first generates multiple, distinct hypotheses explaining why validation failed (e.g., missing section, mis-typed attribute, violated count relation). Each hypothesis then guides the creation of \emph{diverse} candidate fixes rather than minor variations of a single edit. All candidates are revalidated; we select a promising fix and feed that outcome back into the system so it learns which hypotheses and edit strategies resolve specific error classes over time.

\vspace{2pt}
\noindent\textbf{- Operational Loop:} The agentic workflow proceeds through a structured sequence of reasoning and verification stages designed to incrementally improve scenario validity and execution fidelity. Each stage contributes distinct analytical and corrective functions, forming a closed feedback loop between generative reasoning and deterministic validation.

\begin{enumerate}[leftmargin=*, noitemsep]
\item \textbf{Propose:} The agent synthesizes an initial candidate configuration by translating the natural-language specification into a structured CORE XML representation. 

\item \textbf{Validate:} The generated configuration is parsed and evaluated against the CORE schema and associated emulation constraints. Structured validation feedback, including parsing exceptions, attribute mismatches, or rule violations, is collected as formal error signals.

\item \textbf{Hypothesize:} Based on the observed errors, the agent constructs multiple, distinct hypotheses representing alternative explanations for failure (e.g., omitted service blocks, inconsistent routing definitions, or invalid dependency hierarchies). Each hypothesis encodes a potential causal mechanism guiding the next  action.

\item \textbf{Diversify Fixes:} For each hypothesis, the system synthesizes a corresponding corrective variant of the configuration. These variants may involve structural edits (node or link modifications), attribute-level corrections (parameter or type adjustments), or consistency revisions (count or dependency alignment). The diversification stage mitigates convergence toward repetitive or local edits.

\item \textbf{Select:} All candidate revisions are revalidated under identical conditions. The system evaluates performance according to schema compliance, error reduction, and execution viability, selecting the configuration that minimizes residual inconsistencies.

\item \textbf{Learn:} The outcomes of prior iterations are used to update internal preference weights over successful hypothesis–fix pairs.

\end{enumerate}
\noindent

This shifts the process from \emph{error $\rightarrow$ single fix $\rightarrow$ repeat} to \emph{error $\rightarrow$ hypotheses $\rightarrow$ diverse fixes $\rightarrow$ selection $\rightarrow$ learning}, improving robustness under schema drift and reducing stagnation due to deterministic repairs.

\section{Evaluation}
\vspace{-5pt}
We evaluate each model across 100 generated samples. A sample is deemed {correct} if it can be successfully loaded into {CORE} without errors. Rather than making RPC calls, the model uses a {generated XSD schema} to verify structural validity---ensuring compatibility through schema-based validation. The entire process is orchestrated by {Agno AI}, which coordinates the agentic workflow. Additionally, our {agentic framework} is self-correcting: when a discrepancy is detected, the agent autonomously revises its output to meet the schema requirements. For ease of integration, we only focus on existing models and not finetuning, so that users can directly plug-in their LLMs within our agentic frameworks, provide API keys (if they are using proprietary models), and start running immediately.

In Table \ref{tab:zeroshot}, we present the validation performance of a single-shot agent tasked with generating a valid CORE-compliant XML. As shown, the accuracy remains relatively low in this setting. In contrast, Table \ref{tab:gpt_comparison} illustrates the performance of our agentic model, which learns from feedback, analyzes its errors, formulates hypotheses, and iteratively refines its outputs.

\begin{tcolorbox}[learnedbox]
\begin{itemize}
  \item Use only allowed \texttt{pattern} values: \texttt{continuous}, \texttt{periodic}, \texttt{burst}, \texttt{poisson}, \texttt{ramp}.
  \item Conform \texttt{content\_type} to allowed values: \texttt{Random}, \texttt{text}, \texttt{photo}, \texttt{audio}, \texttt{video}, \texttt{gibberish}.
  \item Limit \texttt{segmentation\_kind} to \texttt{Random}, \texttt{Firewall}, \texttt{NAT}, \texttt{CUSTOM}.
  \item Verify \texttt{pattern} values against the specified set: \texttt{continuous}, \texttt{periodic}, \texttt{burst}, \texttt{poisson}, \texttt{ramp}.
  \item Double-check attribute values against their enumerations before submission.
\end{itemize}
\end{tcolorbox}

Based on Tables \ref{tab:gpt_comparison} and \ref{tab:zeroshot}, we observe that single LLM agents do not possess much intrinsic understanding or semantic insight into what constitutes a valid CORE-compliant XML structure. Instead, their performance improvements emerge from pattern learning and rule induction based on feedback. Through repeated iterations, the agents implicitly infer a set of syntactic and structural constraints, such as permissible attribute values, enumeration limits, and field dependencies, without comprehending their underlying meaning. This behavior suggests that the agent’s success arises not from reasoning over schema logic, but from statistical generalization over observed corrections, effectively internalizing a compact rule set that governs valid XML generation. Our aim here is not to prove that one LLM is better than the other, but to observe a general trend, and to also highlight that a powerful open-source model such as gpt-oss is as good as expensive models such as Claude or GPT-4.


\begin{table}[]
\centering
\resizebox{0.9\columnwidth}{!}{%
\renewcommand{\arraystretch}{1.25}
\begin{tabular}{@{}llll@{}}
\toprule
\textbf{\begin{tabular}[c]{@{}l@{}}Evaluation \\ Category\end{tabular}} & \textbf{GPT-4}   & \textbf{\begin{tabular}[c]{@{}l@{}}GPT-OSS \\ (120B)\end{tabular}} & \textbf{Claude}   \\ \midrule
\begin{tabular}[c]{@{}l@{}}Minor \\ Edits\end{tabular}                  & 98.66 $\pm$ 0.94 & 99.33 $\pm$ 0.94                                                   & 95 $\pm$ 2.45     \\ \midrule
\begin{tabular}[c]{@{}l@{}}Moderate \\ Revisions\end{tabular}           & 95.66 $\pm$ 1.69 & 96.66 $\pm$   1.69                                                 & 96.33 $\pm$ 2.05  \\ \midrule
\begin{tabular}[c]{@{}l@{}}Full \\ Regeneration\end{tabular}            & 92.33 $\pm$ 1.69 & 94 $\pm$ 3.55                                                      & 83.66 $\pm$ 11.81 \\ \bottomrule
\end{tabular}%
}
\caption{Comparison of GPT-4, GPT-OSS (120B), and Claude across reference-based change levels. The accuracy reflects the percentage of samples that were successfully executable. 6 rounds and 5 potential fixes (per round) were allowed to generate one scenario.}
\label{tab:gpt_comparison}
\end{table}

\begin{figure}[]
  \centering
  \includegraphics[width=\linewidth]{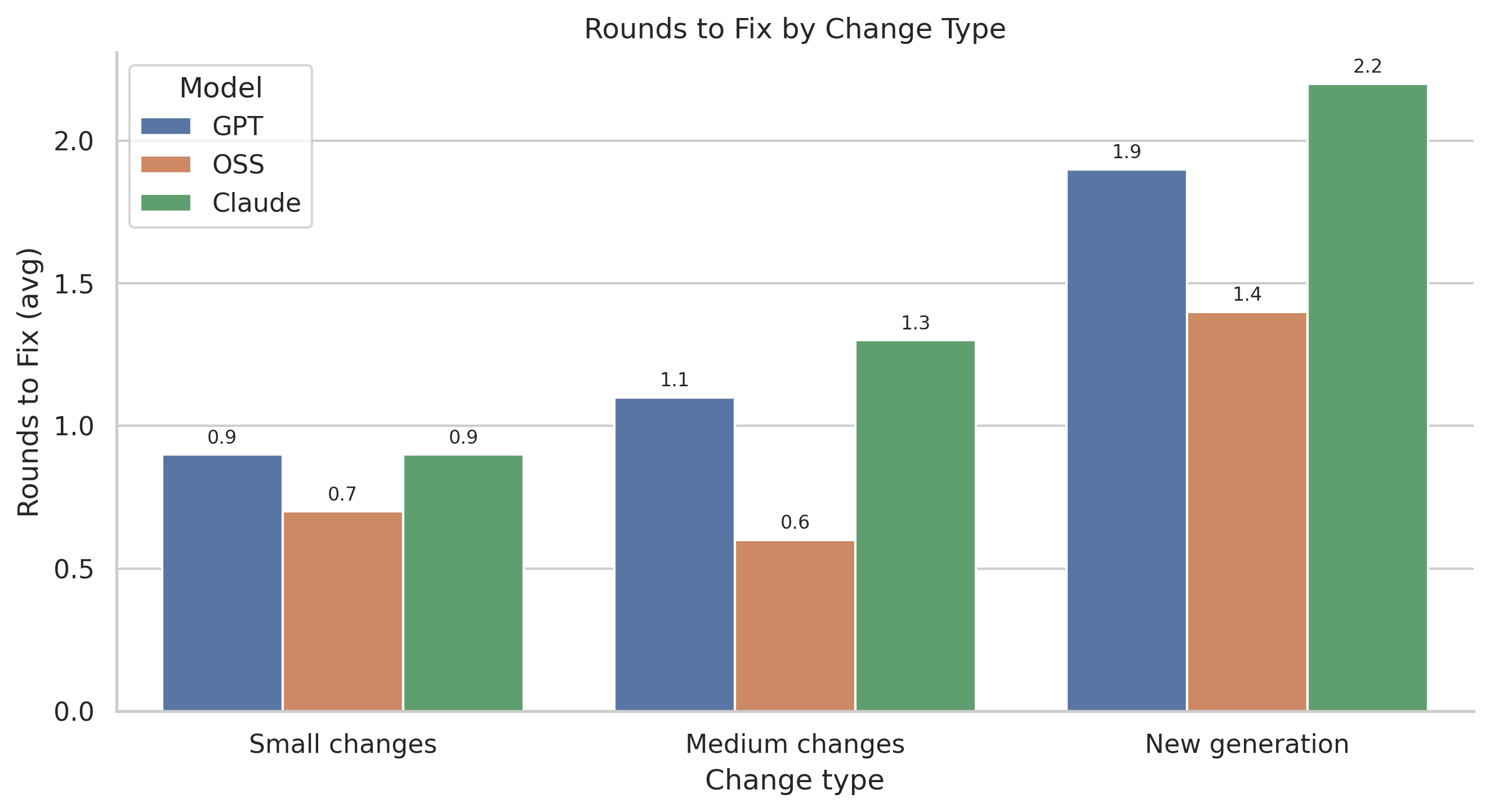}
  \caption{Average rounds taken to fix an XML by change type for GPT, OSS, and Claude.}
  \label{fig:rounds}
\end{figure}

\begin{figure}[]
  \centering
  \includegraphics[width=\linewidth]{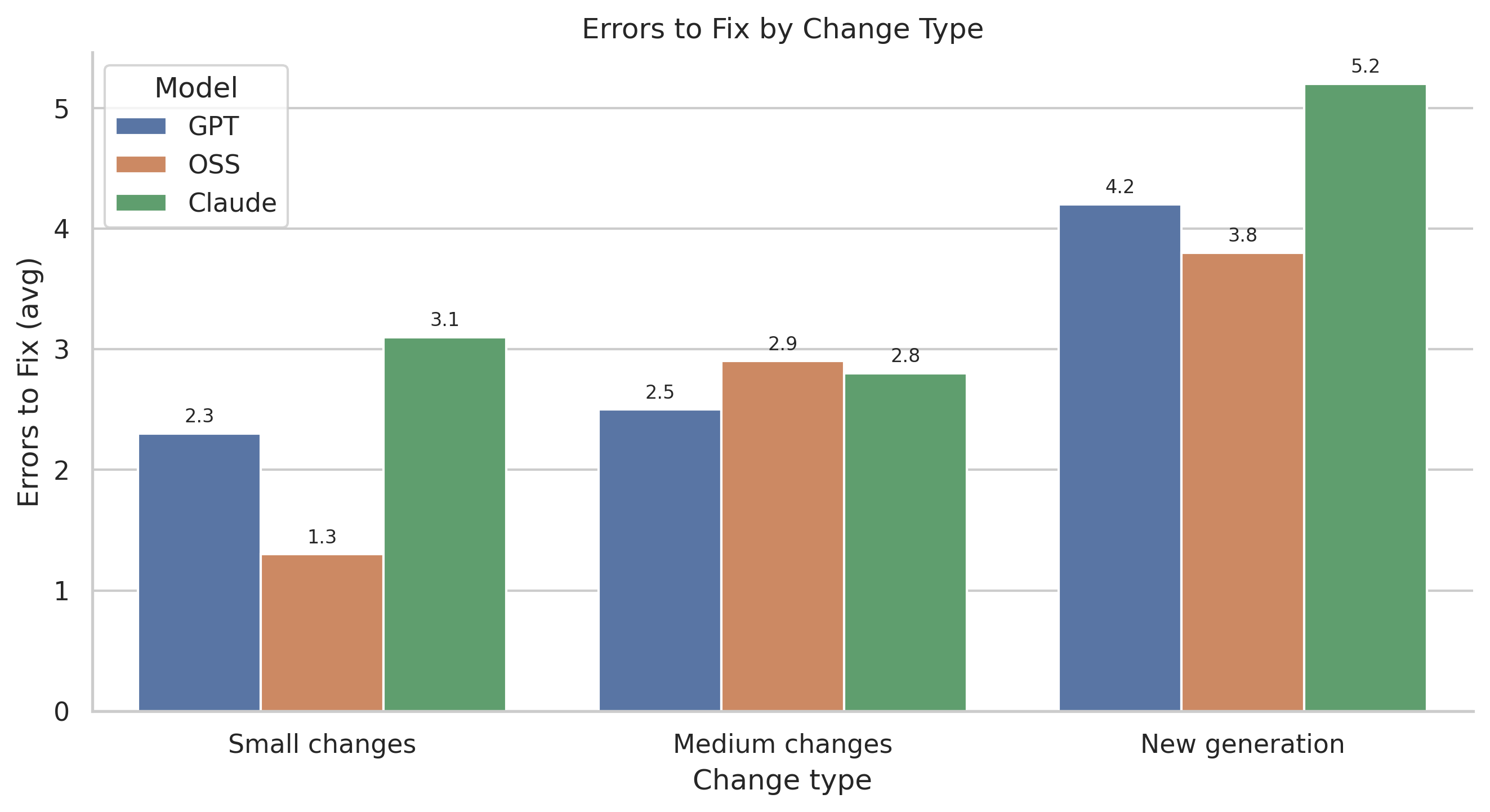}
  \caption{Average errors fixed per XML for each change type for GPT, OSS, and Claude.}
  \label{fig:errors}
\end{figure}

In Figures~\ref{fig:errors} and~\ref{fig:rounds}, we observe that our agentic model effectively understands and resolves patch-related issues. We analyze the agent’s performance along two primary dimensions: (i) the number of rounds and (ii) the number of errors. The \textit{number of rounds} represents the iterations required for the agent to self-correct and produce a CORE XSD-compliant XML. In each round, one or more errors (identified based on predefined violation rules) are detected and fixed. In future iterations, we plan to directly integrate this process with the CORE gRPC interface to enable real-time validation and correction.

\vspace{-5pt}
\section{Conclusion and Future Work}
\vspace{-10pt}
We presented \methodname, a framework combining LLM reasoning with schema-constrained network emulation to automatically generate executable cybersecurity scenarios. By integrating agentic feedback and deterministic validation, the system bridges creative synthesis and operational reliability, reducing manual effort while enhancing realism.
Future work will expand \methodname’s adaptability through dynamic difficulty tuning, real-time validation, and integration of additional domains such as forensics and binary exploitation. We also plan to embed responsible-use safeguards, including controlled vulnerability catalogs and sandboxed deployment. By coupling automation, validation, and ethical design, \methodname establishes a scalable foundation for adaptive cybersecurity training and experimentation aligned with evolving threat landscapes.

\bibliographystyle{IEEEtran}
\bibliography{references}

\appendix

\begin{figure*}[t] 
    \centering
    \includegraphics[width=0.9\textwidth]{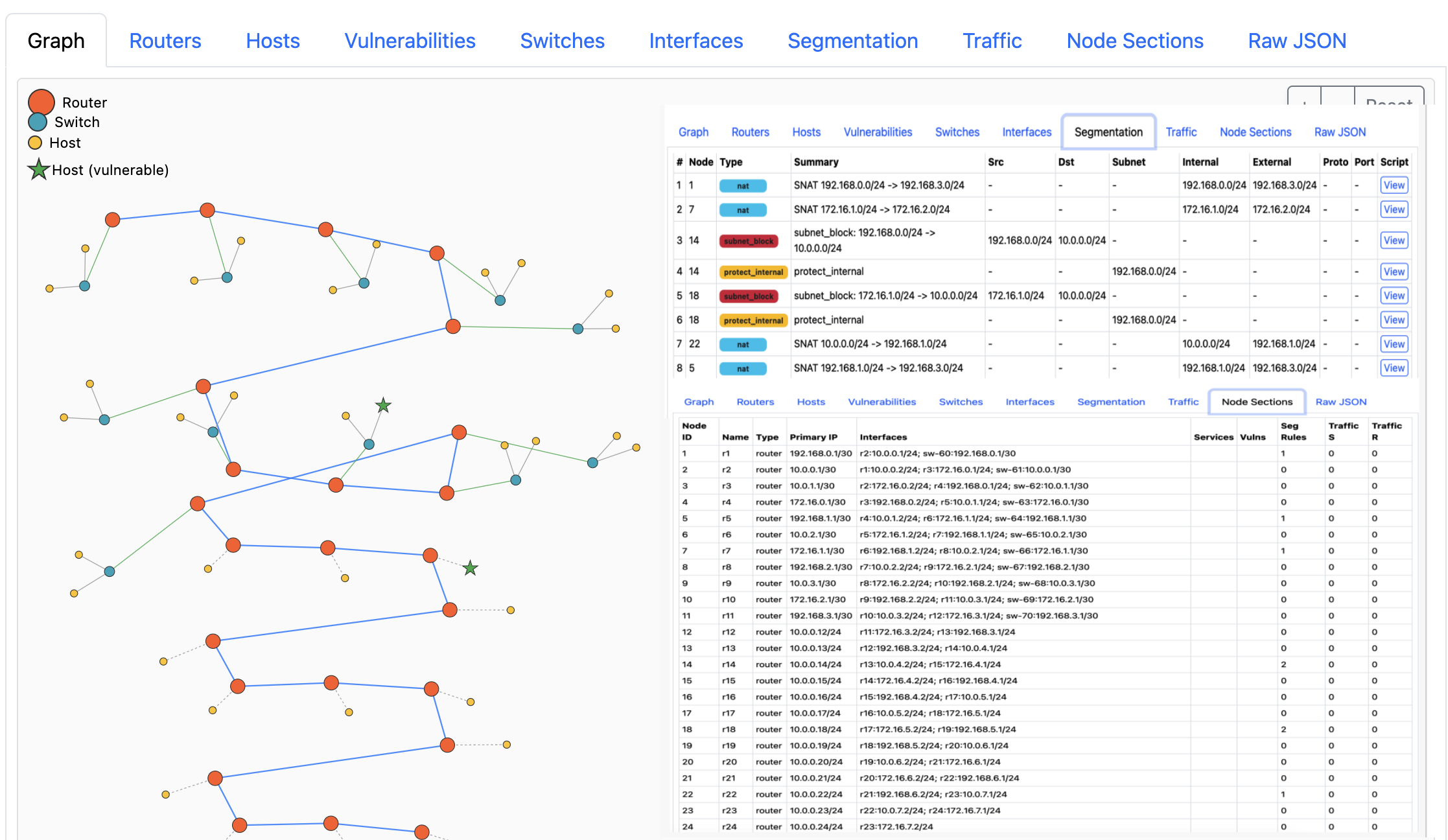} 
    \captionsetup{width=\textwidth,justification=centering,singlelinecheck=false}
    \caption{A generated scenario showing the node graph with routers (large/red), switches (medium blue), and hosts (yellow/small). Vulnerable hosts are green/stars.}
    \label{fig:demo}
\end{figure*}

\subsection{Case Study}
\subsubsection{With Manual Integration:}
To demonstrate some of the capabilities and workflow of \methodname, we developed a network-focused capture-the-flag scenario that includes several vulnerabilities in a realistic scenario. Users must find and exploit the vulnerabilities (that were added to the \methodname vulnerability catalog from \cite{vulhub}) to successfully gain control of a remote machine. By doing so, they will gain access to a \emph{flag} on the remote machine, which will gain them points. The scenario was generated by specifying a scenario with 30 base nodes, sparsely connected, with correctly configured firewalls in place as well as NAT for segregation. A high amount of network traffic should also be included and a total of five remotely exploitable vulnerabilities, including at least one related to SQL injection and one that is specifically, CVE-2022-24706.

As a base scenario, we provided a simple switch and RJ-45 interface (through which a VM could connect). The scenario consisted of a total of 56 nodes; 30 hosts, 24 routers (running RIP), 6 switches, 2 vulnerable nodes, 24 router-to-router edges, 7 traffic pairs, and 6 nodes with segmentation rules (a mix of NAT, subnet-block, and protect-internal: 8 total rules). Additionally, the segmentation rules ensure that any generated traffic can flow between subnetworks; as well as ports required for the SQL (resolved to CVE-2012-2122) and CVE-2022-24706 vulnerable services. Some random services were also introduced, including web (apache2 and nginx) and SSH. The scenario was generated 5 seconds; starting the scenario in CORE took roughly 20 seconds.

To obtain the flags in this generated scenario, the participant will have to first hijack the RIP routes to identify networks that are accessible. Afterward, they will have to use a scan tool such as nmap to enumerate hosts and their services. Digging further into the MySQL server using a tool such as sqlmap would reveal the authentication bypass flaw. Authenticating and then accessing database table content would reveal the flag. The SSH server is exploited by finding a binary on this server. Similarly, the CouchDB vulnerability (which allows null-authentication and privilege escalation) could be identified using nmap due to an \emph{accept} rule in the firewall and then exploited using a tool like metasploit. Figure \ref{fig:demo} shows some of the results from the scenario generation. While we focused on network-level vulnerabilities, this work can easily expand to cover a wide range of domains, including binary analysis, cryptography, encoding, forensics, and many others.

\subsubsection{With Agentic Integration:}
\begin{figure*}[]
  \centering
  \vspace{2mm} 
  \includegraphics[width=0.9\textwidth]{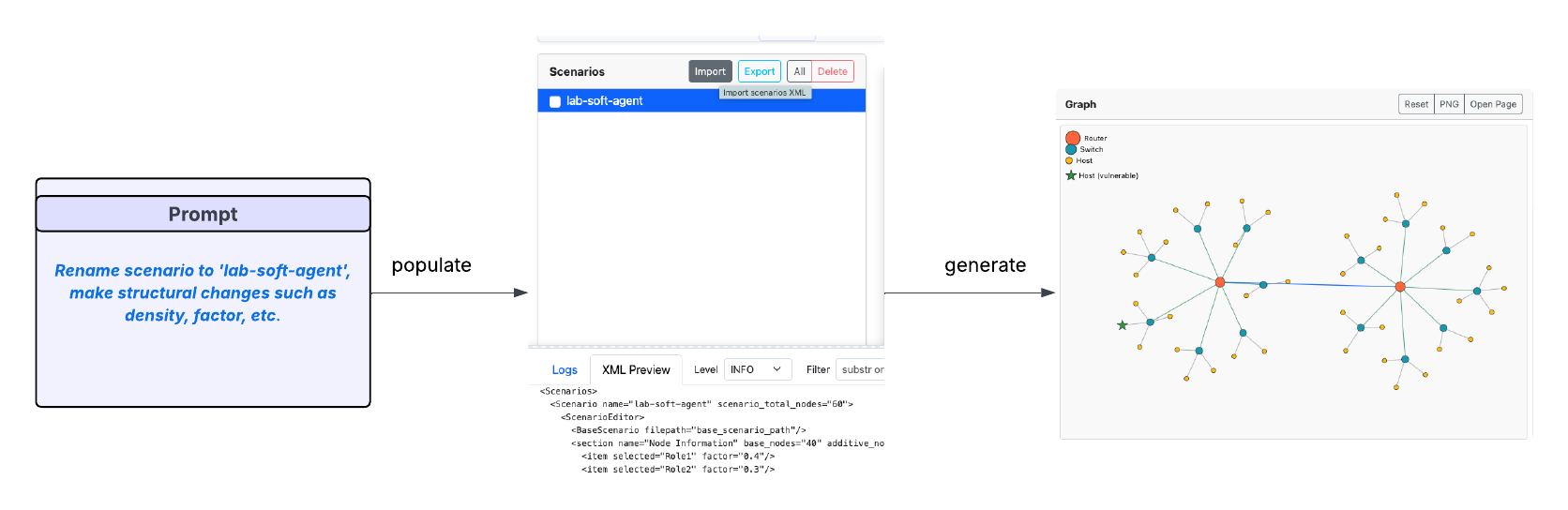}
  \caption{CTF generation pipeline (case (i): small–medium changes). 
  The reference \texttt{.xml} is ingested by TopoGen, integrated with Vulhub for CTI-derived vulnerability introduction, 
  and the resulting scenario includes network connections, routers, devices, and a vulnerable host.}
  \label{fig:ctf-gen}
  \vspace{-3mm} 
\end{figure*}
In Fig.~\ref{fig:ctf-gen} we show how our agentic system integrates with the tool developed in \methodname. Our agentic framework supports two operational modes: (i) \textbf{edit-from-reference}, where the agent takes a reference \texttt{.xml} topology and produces a variant with small–medium edits (e.g., host count, subnet ranges, routing policy, service mix, or vulnerability selection); and (ii) \textbf{generate-from-scratch}, where the agent synthesizes a new \texttt{.xml} that loads without manual fixes via an internal self-correction loop (schema checks, consistency constraints, and a dry-run import). The figure instantiates mode~(i): the agent (1) parses the reference \texttt{.xml} and proposes a bounded change-set; (2) patches the \texttt{.xml} while enforcing topology constraints (unique names, link endpoints, address allocation); (3) validates the result with schema and simulator checks; and (4) loads the artifact into \methodname, which materializes the logical graph into a runnable scenario. \methodname is integrated with \textbf{Vulhub} to introduce CTI-derived vulnerabilities (selected services/CVEs bound to specific hosts). The final scenario contains the concrete network (routers, links, devices) and at least one vulnerable host reachable along the intended attack path, making the environment ready for CTF-style evaluation and automated scoring.


\end{document}